\documentclass{elsart}
\usepackage{ifpdf}
\usepackage{graphicx,amssymb,lineno}
\usepackage{amssymb,lineno,amsfonts}
\usepackage{graphicx}   % Include figure files
\usepackage{dcolumn}    % Align table columns on decimal point
\usepackage{bm}         % bold math
\usepackage{bbm}
\usepackage{mathrsfs}
\usepackage{upgreek}
\usepackage{mathtools}

\begin{document}
\begin{frontmatter}
\title{Measurement of Translational Diffusion Constant using Noon State}

\author{Abhishek Shukla,}
\author{Manvendra Sharma, and}
\author{T. S. Mahesh}$^*$
\ead{*mahesh.ts@iiserpune.ac.in}
\address{Department of Physics and NMR Research Center, \\
				 Indian Institute of Science Education and Research,
				 Pune 411 008.}
\date{\today}

\begin{abstract}
{
A method for measuring translational diffusion constant in liquids 
via NOON state is described using a quantum
circuit and is experimentally demonstrated using a model system.  
When compared with the standard single quantum method,
the NOON state method requires shorter diffusion delays and weaker
gradients.  These improvements depend on the on the size of the NOON state.
Due to the high sensitivity of the NOON state for the changes in the local 
magnetic fields, this method enables studying slow diffusion
and studying diffusion with limited strengths of pulsed-field-gradients.
}
\end{abstract}

\end{frontmatter}

\section{Introduction}
Driven by the internal thermal energy, the atoms or molecues of a bulk
matter exhibit random translational motion, which is termed as translational
diffusion \cite{ghez,march}.  Diffusion explains the net flux of particles moving from a 
region of higher concentration to the lower concentration, although
diffusion process exists even when there is no macroscopic concentration 
gradient.  Diffusion is a fundmental transport mechanism in liquids and gasses and therefore
measurement of diffusion constant is important for understanding 
many physical, chemical, and biochemical processes \cite{Meerwall,waldeck,Zoran,liposome}.  
The diffusion constant ($D$) is described as the amount 
of a particular substance that diffuses across a unit area in unit time 
under the influence of a unit concentration gradient \cite{ghez}.  
According to the
Stokes-Einstein theory, for a fluid with visocity $\eta$ at a temperature $T$,
the diffusion constant is given by
\begin{eqnarray}
D = \frac{kT}{6 \pi \eta r_s},
\end{eqnarray}
where $k$ is the Boltzman constant and $r_s$ is the Stokes radius of the 
diffusing particle \cite{einstein,Tyrell}.  The denominator in the above expression is termed as
the friction coefficient.

\begin{figure}
\begin{center}
\includegraphics[width=8cm]{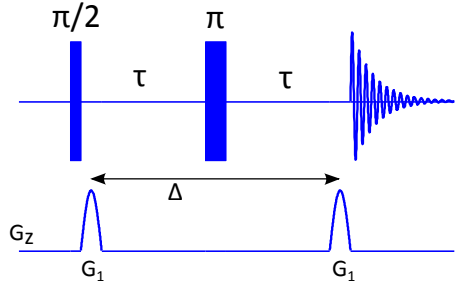}
\caption{Hahn-echo pulse sequence with two equal PFGs 
for the measurement of diffusion constant.}
\label{hahndiff}
\end{center}
\end{figure}
Diffusion constant of a liquid can be measured by NMR either 
with the help of relaxation studies or more conveniently using 
pulsed-field-gradients (PFGs)
\cite{cpdiff}.   Fig. \ref{hahndiff} shows a standard pulse
sequence based on the Hahn-echo method for measuring diffusion constant \cite{stejskal,price2}. 
In this method,
maximum amplitude of Hahn-echo can be obtained by a refocussing 
$\pi$ pulse placed between two PFGs if the molecues do not change their 
place during the period $\Delta$.  
Diffusion renders the molecules to change their place and 
therefore for a fixed duration $\Delta$, the echo amplitude decreases with 
the strength of the PFGs.  
Such PFG methods are widely used for the measurement of diffusion constants
and already numerous improved sequences are available \cite{pricea,priceb,johnson}.
For example, replacing Hahn-echo by stimulated-echo sequence reduces the
decay due to transverse relaxation and results in enhanced echo amplitudes \cite{Hahn,Tanner2}.
Application of long-lived singlet states allows one to study slow diffusion
requiring long intervals ($\Delta$) between the PFGs \cite{bodenhausen}.  
Fast single-scan measurement of 
diffusion has also been demonstrated by realizing 
z-coordinate dependent PFG strengths
using frequency swept refocusing pulses \cite{keeler}.

In this letter we describe another approach inspired by quantum information
theory to speed-up the measurement of diffusion constant.  We propose to 
prepare the diffusing spin system into a NOON state.  Although this
method has similarities with the multiple-quantum method for studying diffusion 
\cite{kay,chapman,zhang},
the current method differs in actual implementation and its scope.
We briefly outline the theory in section 3 and we describe the 
experimental demonstration in section 4.

\section{Diffusion constant by NOON state}
\subsection{NOON state}
We shall follow the notations of quantum information theory where $\pm 1/2$
states of a spin 1/2 nuclues is represented by basis kets $\vert 0 \rangle$ and
$\vert 1 \rangle$ of a quantum bit or a `qubit'.  
The NOON state of an $N$-qubit system is a superposition
%state
%\begin{eqnarray}
%\vert MSSM \rangle = (\vert M, S \rangle + \vert S, M \rangle)/\sqrt{2},
%\end{eqnarray}
% where $M$ and $S$ represent
%the number of qubits in states 0 and 1 respectively.  
%The above state is often referred to as `many-some,some-many' states \cite{jonesscience}.
%The NOON state is a special form of $\vert MSSM \rangle$ state that is 
%maximally entangled and is 
%a superposition 
of all the $N$ qubits
being in state 0 with all those in state 1:
\begin{eqnarray}
\vert N00N \rangle &=& (\vert N, 0 \rangle + \vert 0, N \rangle)/\sqrt{2} \nonumber \\
                   &=& (\vert 00 \cdots 0 \rangle + \vert 11 \cdots 1 \rangle)/\sqrt{2}
\end{eqnarray}
\cite{boto}.
Recently several applications of NOON states have been discovered \cite{lee,dowling,chen}.
It has also been used for sensing weak magnetic fields \cite{jonesscience}.

\begin{figure}
\begin{center}
\includegraphics[width=8cm]{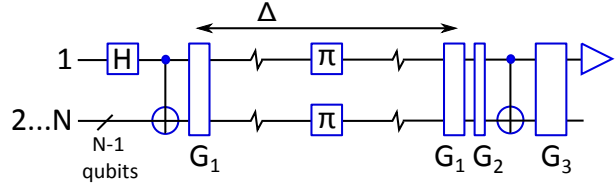}
\caption{Circuit for the measurement of diffusion constant
using NOON state.  NOON state is initialized using a Hadamard
gate (H) and a CNOT ($\oplus$) gate.  The diffusion of particles
is monitored using two identical PFGs of strength G$_1$ placed on either side of
$\pi$ pulses.  The second CNOT converts the evolved NOON state into 
an observable single quantum coherence of the 1st qubit.  The signal of
the first qubit is measured after the
PFGs G$_2$ and G$_3$ select the coherence transfered from
the NOON state.}
\label{nooncircuit}
\end{center}
\end{figure}

The circuit for the preparation of
NOON state, shown in the first part of Fig. \ref{nooncircuit},
consists of a Hadamard gate (H) and a CNOT gate.  This circuit acts on a 
quantum register with a single `control' qubit and a set of $(N-1)$ `target'
qubits initialized in $\vert 00 \cdots 0 \rangle$ pure state
\begin{eqnarray}
\vert 00 \cdots 0\rangle 
&\stackrel{\mathrm{H_{control}}}{\longrightarrow}& 
\frac{1}{\sqrt{2}}(\vert 0 \rangle + \vert 1 \rangle) \vert 00 \cdots 0 \rangle
\nonumber \\
&\stackrel{\mathrm{CNOT}}{\longrightarrow}&
\frac{1}{\sqrt{2}}(\vert 00 \cdots 0 \rangle + \vert 11 \cdots 1 \rangle)
= \vert N00N \rangle.
\end{eqnarray}

Preparing a pure NOON state is out of bounds
from a highly mixed thermal equilibrium state that is available in
liquid state NMR systems at ordinary temperatures.
However, it is rather easy to prepare a pseudopure NOON state
which is of the form
\begin{eqnarray}
\rho_{N00N} = \frac{1-\epsilon}{2^N}\mathbbm{1} + \epsilon \vert N00N \rangle \langle N00N \vert,
\label{rhonoon}
\end{eqnarray}
where the scalar quantity $\epsilon$ is a measure of purity, 
which is of the order of $10^{-5}$ in typical NMR setups \cite{corypnas}.
The identity part does not evolve under unitary dynamics neither
does it give raise to NMR signals.  Therefore we ignore this part
as a background and continue using ket notation $\vert N00N \rangle$
for simplicity.

The CNOT implementation is particularly easier with a heteronuclear spin
system of type AM$_{N-1}$, wherein each M spin of the 
$N-1$ magnetically equivalent spins,
is having an indirect spin-spin interaction with the common heteronuclear
spin A with a coupling constant of, say $J$ Hz.  We shall
use the single heteronucles as the `control' qubit and the rest as the
`target'.
Such a system allows parallel implementation of CNOT gate 
(upto a global phase)
using the sequence
\begin{eqnarray}
U_\mathrm{CNOT} = 
Y_\mathrm{A}^2 
\bar{Y}_\mathrm{M} 
U_J 
Y_\mathrm{A,M}^2 
U_J 
\bar{Y}_\mathrm{A} 
\bar{X}_\mathrm{A} 
Y_\mathrm{A} 
\bar{Y}_\mathrm{M} 
\bar{X}_\mathrm{M}
%Uym2*UY1*UJ*UY1*UY2*UJ*Uym1*Uxm1*Uy1*Uym2*Uxm2
\end{eqnarray}
\cite{cory2,jones1}.
Here $X = e^{-i(\pi/2)I_x}$ and $\bar{X} = e^{+i(\pi/2)I_x}$
are rotations on the spins indicated by their subscripts ($I$ 
is the spin operator and $Y$, $\bar{Y}$ are defined similarly). 
Under
$U_J = e^{-i (\pi/2) J I_z^A \sum_\mathrm{M} I_z^\mathrm{M} }$ all the target qubits simultaneously
undergo J-evolution with the control and is realized simply by a 1/(4J) delay.
In our experiments we have replaced the Hadamard gate on A with a pseudo-Hadamard
gate which can be implemented by a single $Y_A$ pulse.  The rest of the circuit in
Fig. \ref{nooncircuit} is described in the following.

\subsection{Phase encoding and diffusion measurement}
Consider first the simple Hahn-echo sequence for measuring
the diffusion constant as illustrated in Fig. \ref{hahndiff}.
For simplicity, we ignore decoherence and other experimental
errors in the following analysis.
The first gradient encodes the longitudinal distribution of the
sample such that a spin in position $z$ gets a phase 
$\phi(z) = \gamma z G_z \delta$, where $\gamma$ is the gyromagnetic
ratio of the spin and $\delta$ is the effective duration (duration
$\times$ shape-factor) of the gradient $G_z$.  Under this gradient
the pseudopure coherence $\vert 0 \rangle + \vert 1 \rangle$
(ignoring the backgroud)
evolves to $\vert 0 \rangle + e^{i\phi} \vert 1 \rangle$.
After the central $\pi$ pulse flips the basis states the second gradient
encodes another identical phase so that the resulting state becomes
$e^{i\phi}\vert 1 \rangle + e^{i\phi} \vert 0 \rangle
\equiv \vert 0 \rangle + \vert 1 \rangle$, up to a global phase.  Thus if the
spin does not change the position there is no change in the relative
phase.  However if, during the period $\Delta$, 
the spin changes the position by an effective amount $dz$,
the state after the second gradient will be
$\vert 0 \rangle + e^{id\phi} \vert 1 \rangle$, where
the relative phase is $d\phi = \gamma dz G_z \delta$.

Due to the random molecular motion of a large number of spins each of which
acquiring a different relative phase, the average Hahn-echo does not
acquire a net phase but diminishes in amplitude.  Theoretically, the
decay of the Hahn-echo signal $S$ can be given as,
\begin{eqnarray}
S(G_z) = S_0 \exp\left\{-\gamma^2 G_z^2 \delta^2 D (\Delta  - \delta/3 )\right\},
\label{sgz}
\end{eqnarray}
where $D$ is the diffusion constant, $S_0 = S(0)$ is the normalization
factor, and $\Delta - \delta/3$ is the correction to $\Delta$ 
due to the finite durations $\delta$ of the PFGs \cite{pricea}.

Similar analysis can now be carried out for the NOON state using the
circuit shown in Fig. \ref{nooncircuit}.  Under the $G_1-\pi-G_1$
sequence, a spin system in NOON state diffusing through a distance
$dz$ acquires a net relative phase and becomes
$(\vert 00 \cdots 0 \rangle + e^{id\phi} \vert 11 \cdots 1\rangle)/\sqrt{2}$.
The relative phase acquired is $d\phi = \gamma_\mathrm{eff} dz G_z \delta$,
where $\gamma_\mathrm{eff} = \{1+(N-1)\gamma_M/\gamma_A\}\gamma_A = l\gamma_A$.
The factor $l$ is also known as `loopsidedness' of the NOON state. Larger the value
of $l$, more sensitive is the NOON state for the phase encoding.  
Since NOON state is a multiple quantum coherence it is necessary to convert
it back to single quantum coherence before detection.  This conversion can
efficiently be carried out using a second CNOT gate:
\begin{eqnarray}
\frac{1}{\sqrt{2}}(\vert 00 \cdots 0 \rangle + e^{id\phi} \vert 11 \cdots 1\rangle) \stackrel{\mathrm{CNOT}}{\longrightarrow} \frac{1}{\sqrt{2}} 
(\vert 0 \rangle + e^{id\phi}\vert 1 \rangle) \vert 00 \cdots 0 \rangle.
\end{eqnarray}
Thus the phase encoding due to the diffusion, i.e., $d\phi$ 
has been transfered to one transition of the control spin.  However as explained
before, for a large number of molecules undergoing diffusion, the above phase
encoding results in the attenuation of the control transition.
Instead of starting with an initial pseudopure state,
%$\vert 00 \cdots 0 \rangle$ state, 
%only a single transition will be seen after
%the second CNOT.  However, 
it is rather convenient to start with a thermal
equilibrium state and use two PFGs G$_2$ and G$_3$ to select out the desired
coherence pathway that is passing through the NOON state.  The ratio of these
two PFGs are adjusted depending on the relative gyromagnetic ratio: 
$G_3 = G_2((N-1)\gamma_M/\gamma_A+1)$.

In the case of large loopsidedness $l$, the NOON state helps us to
study diffusion with weaker PFGs and smaller durations ($\Delta$) 
between them.  An experimental demonstration of this method for a model system
is described in the following section.

\section{Experiment}
\begin{figure}
\begin{center}
\includegraphics[width=7cm]{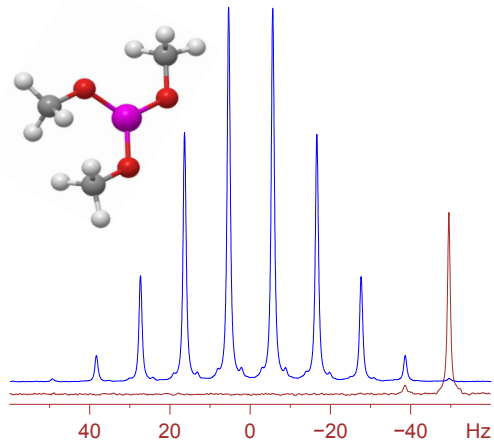}
\caption{The $^{31}$P spectra corresponding to single quantum excitation
from thermal equilibrium (upper trace) and after converting the NOON state into the single 
quantum coherence using a CNOT gate (lower trace). 
The inset 
displays the molecular structure of trimethylphosphite (P(OCH$_3$)$_3$).
}
\label{diffspec}
\end{center}
\end{figure}

The sample consisted of 100 $\upmu$l of trimethylphosphite 
(see inset of Fig. \ref{diffspec})
dissolved in 500 $\upmu$l of dimethyl sulphoxide-D6.  
All the experiments were carried out on a 500 MHz Bruker
NMR spectrometer at an ambient temperature of 300 K.
Each of the nine magnetically equivalent $^1$H spins are coupled to
the $^{31}$P spin via indirect spin-spin interaction with a coupling 
constant of $J=11$ Hz thus forming an AM$_9$ spin system.
Accordingly phosphorous spectrum splits into 10 lines as shown
in Fig. \ref{diffspec}.  An initial INEPT transfer was used to
enhance the $^{31}$P polarization.  Then, 
as described in the previous section, 
NOON state was prepared with a pseudo-Hadamard gate and a subsequent CNOT gate,
and is converted
back to a single quantum coherence using a second CNOT. 
%The PFGs G$_2$ and
%G$_3$ are adjusted to select the 10-quantum coherence pathway. 
The selection PFGs were adjusted such that G$_3$/G$_2$ = $9\gamma_H/\gamma_P+1$
= $23.23$, to select the 10-quantum coherence pathway. 
A two-step phase cycle of $\bar{X}_\mathrm{M}$ pulse of the first
CNOT along with the receiver phase helped to reduce artifact signals.
The single transition in $^{31}$P spectrum of Fig. \ref{diffspec} indicates
the selection of NOON-state coherence pathway.

\begin{figure}
\begin{center}
\includegraphics[width=8cm]{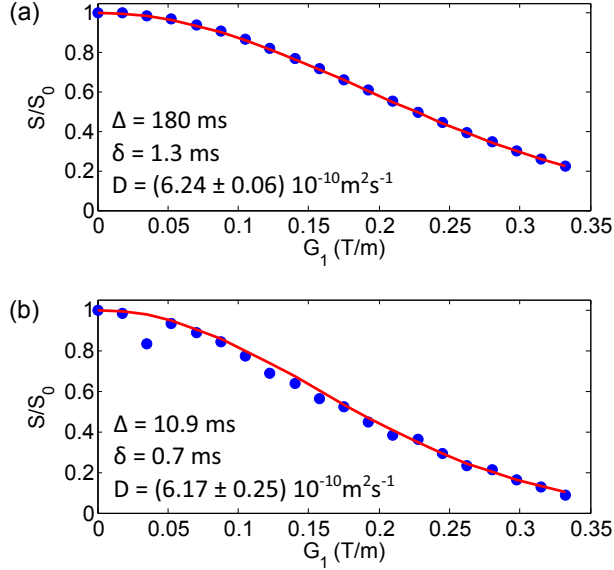}
\caption{Intensity of the echo signals as a function of the gradient strength G$_1$
with (a) $^1$H single quantum coherence and (b) NOON state.  The dots represent the 
experimental data and the lines represent the Gaussian fit.  The insets in both
the figures display the parameters of the experiments.
}
\label{expt}
\end{center}
\end{figure}

We first carried out the single quantum diffusion experiment as described 
in Fig. \ref{hahndiff}.  The parameters of the experiment are shown in
Fig. \ref{expt}a. While keeping all other parameters fixed, we varied only the
strength of the PFGs G$_1$ in the range 0 to 0.3325 T/m in 20 equal intervals.
The integrated intensity of the $^1$H signal as a function of G$_1$ is
shown in Fig. \ref{expt}a. The diffusion constant, 
$D = (6.24 \pm 0.06) 10^{-10}$ m$^2$ s$^{-1}$
was obtained by fitting these data points to the Gaussian expression given in 
(\ref{sgz}).

Then we prepared the NOON state as explained above, and diffusion experiment
was carried out as in Fig. \ref{nooncircuit}.
The main parameters of the experiment are shown in Fig. \ref{expt}b. 
The lopsidedness for this system is $l=9.4$ and accordingly 
the NOON state experiment requires much shorter $\Delta$ and $\delta$ values
to reach similar attenuation of the echo signal as that of the single
quantum experiment.  In Fig.\ref{expt}b, the integrated intensity of the single transition,
after the selection of NOON-state coherence pathway, is plotted against G$_1$.  The Gaussian fit, again
using the expression (\ref{sgz}), with $\gamma_\mathrm{eff} = l \gamma_H$ lead to the
diffusion constant $D = (6.17 \pm 0.25) 10^{-10}$ m$^2$ s$^{-1}$.  While the two
methods gave identical values for the diffusion constant, the error-bar is slightly 
larger in the latter case
due to the added complexity of the circuit, the imperfections in the selection
of coherence pathway, reduced signal to noise, and due to the faster relaxation
of the NOON state.  On the otherhand, the shorter timescales of the NOON-state
experiment suggests its possible applications in studying slow diffusion or
studying diffusion with limited PFG strengths.  

\section{Conclusions}
We described the application of NOON states in studying translational
diffusion in liquids.  Although the basic principle is general, the method is 
particularly convenient in the presence of a set of magnetically equivalent 
nuclei interacting with a heteronucleus via J-coupling.  We have demonstrated
the experimental measurement of diffusion constant in trimethylphosphite, 
which is a $AM_9$ spin system.  Both single-quantum and the NOON-state 
experiments lead to identical values for the diffusion constant, but
the error-bar was slightly larger in the latter case due to the additional
complexities.  However, the NOON state experiment required weaker pair
of PFGs and an order of magnitude shorter duration between them, indicating the possible applications
in studying slow diffusion.  The other available method for measurement of
slow diffusion involves the use of long-lived singlet states \cite{bodenhausen}, which requires
a pair of weakly interacting homonuclear spins spatially isolated from the 
rest of the spins \cite{levittsinglet}.
In the current method we prepare a state that is highly sensitive to the
gradient encoding of the z-coordinate and hence is able to capture the 
effects of diffusion even in shorter time-scales and with weaker PFGs.
It might also be possible to combine the NOON state technique with the
single scan techniques to achieve ultra-fast diffusion measurements.  Such
an experiment will then allow studying dynamic situations with time-varying
diffusion constants.

\section*{Acknowledgment}
Authors gratefully acknowledge Prof. Anil Kumar of IISc-Bangalore,
S. S. Roy of IISER-Pune, and Dr. T. G. Ajithkumar of NCL-Pune for  discussions.

\end{document}